\documentclass[twocolumn,showpacs,amsmath,amssymb,pr,superscriptaddress]{revtex4-1}
\usepackage{graphicx}
\usepackage{epsf}
\usepackage{ulem}
\usepackage{color}
\usepackage[unicode=true,colorlinks=true,urlcolor=blue,citecolor=blue]{hyperref}

\newcommand{\cred}[1]{{\color{black} #1}}
\newcommand{\ild}[1]{{\color{black} #1}}
\newcommand{\ymg}[1]{{\color{black} #1}}

\begin{document}

\title{Composite fermions in a wide quantum well in the vicinity of the filling factor 1/2}

\author{I. L. Drichko}
\author{I. Yu. Smirnov}
\affiliation{Ioffe Physical–Technical Institute, Russian Academy of Sciences, St. Petersburg, 194021 Russia}
\author{A.~V.~Suslov}
\affiliation{National High Magnetic Field Laboratory, Tallahassee, FL 32310, USA}
\author{D.~Kamburov}
\author{K.~W.~Baldwin}
\author{L.~N.~Pfeiffer}
\author{K.~W.~West}
\affiliation{Department of Electrical Engineering, Princeton University, Princeton, NJ 08544, USA}
\author{Y. M. Galperin}
\affiliation{Ioffe Physical–Technical Institute, Russian Academy of Sciences, St. Petersburg, 194021 Russia}
\affiliation{Department of Physics, University of Oslo, PO Box 1048 Blindern, 0316 Oslo, Norway}

\begin{abstract}

Using acoustic method we study dependences of
transverse
AC conductance, $\sigma (\omega)$, on magnetic field, temperature  and  the amplitude of AC electric field in a wide (75~nm)  quantum well (QW) structure focusing on the vicinity of the filling factor $\nu =1/2$. Measurements are performed in the frequency domain 30-307~MHz and in the temperature domain 20-500~mK. Usually, in wide QW structures closely to  $\nu =1/2$ the fractional quantum Hall effect (FQHE) \ild{regime is realized at some parameters of the sample. However, in our structure, \ild{at $\nu =1/2$} it is a compressible state corresponding to gas of composite fermions which is observed}. This is confirmed by apparent frequency independence and weakly decreasing temperature dependence of $\mathrm{Re}\, \sigma(\omega)$. Comparing the dependences of this quantity on temperature and  power of the acoustic wave we conclude that the observed nonlinear behavior of the conductance is compatible with heating of the composite fermions by the acoustic wave. For comparison, we also study the vicinity of $\nu = 3/2$ where the FQHE regime is clearly observed.

\end{abstract}

\maketitle

\section{Introduction}

The fractional quantum Hall effect (FQHE)~\cite{Tsui_PhysRevLett.48.1559} has   become the focus of considerable
theoretical and experimental attention mainly because of deeply non-trivial physics.  Recently, the interest to  this phenomenon was boosted by its potential application in topological quantum computing~\cite{Nayak_RevModPhys.80.1083}.
Despite numerous experimental efforts during the past two decades, however, an understanding of its origin
remains incomplete.

The FQHE is mainly seen in high-quality two-dimensional (2D) electron systems in the lowest ($N = 0$) Landau level at odd-denominator
fillings~$\nu$~\cite{Jain_2007}. In the first Landau level the FQHE exists at $\nu=5/2$~\cite{Willett_PhysRevLett.59.1776,Pan_PhysRevLett.83.3530}.

The possibility of an even-denominator FQHE in the lowest Landau level has been predicted theoretically and considered in many publications, while experimentally, FQHE states
at $\nu= 1/2$ have been seen in electron systems confined to either double-layer~\cite{Eisenstein_PhysRevLett.68.1383,Suen_PhysRevLett.68.1379} or wide GaAs quantum well (QW)
systems~\cite{Suen_PhysRevLett.72.3405}, see Ref.~\cite{Shabani_PhysRevB.88.245413} for a review. In standard, relatively narrow QWs close to the filling factor of 1/2 the FQHE is not observed, and magnetic field dependence of the conductance can be rather described as a gas of so-called \textit{composite fermions} (CFs)~\cite{Jain_2007,Willett_PhysRevB.47.7344}. This was confirmed also in the experiments on so-called geometric resonance of CFs~\cite{Mueed_PhysRevLett.114.236406}, which turns out to exist both in compressible and incompressible state.

The electron states in low-dimensional electron gas are extremely sensitive to the parameters of the structure, especially in wide QWs where electron-electron repulsion lifts the potential energy near the well center and creates an effective barrier. At high densities the charge distribution can behave as ``bilayerlike''. Nevertheless, the interlayer tunneling, quantified by the
symmetric-to-antisymmetric subband separation, $\Delta_{SAS}$, can be substantial.  To classify the electronic ground state one has to compare this quantity with intralayer Coulomb energy $E_C \equiv e^2/4\pi \epsilon l_B$ where $l_B=(\hbar c/eB)^{1/2}$ is the magnetic length while $\epsilon$ is the dielectric constant. Another important quantity is the strength of the interlayer Coulomb interaction, $E_{i}\equiv e^2/4\pi \epsilon d$. where $d$ it the interlayer distance. Regions characterized by different relationships between $\Delta_{SAS}$, $E_C$ and $E_i$  constitute a set of ``phase diagrams'', ground states of the electronic system being different in different regions. Recently, the phase diagrams for $\nu$ close to 1/2 were analyzed in~\cite{Shabani_PhysRevB.88.245413} based on experimental findings and compared with theoretical predictions of~\cite{Peterson_PhysRevB.81.165304}. It is concluded that while there is some overall qualitative agreement, there are also significant quantitative discrepancies. Phase diagrams in tilted magnetic fields were analyzed in~\cite{Hasdemir_PhysRevB.91.045113}.

In the present paper, we report on measurements of AC conductance, both linear and nonlinear in amplitude, in a wide QW structure at different temperatures, frequencies and amplitudes of AC electric field close to the filling factor $\nu=1/2$ ($l_B = 0.7\times 10^{-6}$~cm).
The behavior observed is compatible with response of a compressible state -- gas of composite fermions, rather than with the FQHE.

\section{Experimental details}

\paragraph{Sample} --
The  samples were high-quality multilayered $n$-type GaAlAs/GaAs/GaAlAs structures with a wide ($W=75$~nm) quantum well (QW).
The GaAs QW was Si $\delta$-doped on both sides. The wide QW and the Si $\delta$-layer were separated by 820~${\AA}$ AlGaAs spacers. The structure was grown on GaAs substrate with (001) growth direction and was covered by 10~nm n-GaAs cap. Overall the QW located at the depth $d_0=197$~nm below the surface of the sample.  After illumination of the sample with infrared light with a light emitting diode at low temperatures (down to 15 K) the electron density $n_e$ was $3\times 10^{11}$~cm$^{-2}$ and the mobility $\mu$ was $2.4\times 10^{7}$~cm$^2$/Vs at $T=0.3$~K.
\ild{The electron density initially was specified by the growth parameters and then was confirmed by the value of the period of acoustic quantum oscillations at $B <$2~T.}
Distribution of the electron density and structure parameters $\Delta_{SAS}=0.57$~meV and $d=56$~nm were calculated in Refs.~\cite{Shchurova,Drichko_PhysRevB.97.075427}.
The experimental value of $\Delta_{SAS}$ is 0.42~meV~\cite{Drichko_PhysRevB.97.075427}.
%

\paragraph{Method} --
We use the standard acoustic method, see, e.g.,~\cite{Drichko_PhysRevB.62.7470}.
A traveling surface acoustic wave (SAW) excited by an  inter-digital transducer propagated along a surface of a piezoelectric crystal LiNbO$_3$ while the sample was slightly pressed onto this surface. The interaction of the SAW electric field with charge carriers in quantum well, the SAW attenuation $\Gamma$  and its velocity shift $\Delta V/V$ are governed by the complex AC conductance, $\sigma (\omega) \equiv  \sigma_1 (\omega) - i  \sigma_2 (\omega)$.
The SAW technique has shown to be very efficient tool as it provides a probe-less way  to study both real (1) and imaginary (2) components of the AC conductance.

We measured $\Gamma$ and $\Delta V/V$ within the temperature domain of 20--500~mK, the frequency domain of 30--307~MHz, and in perpendicular magnetic fields up to 18~T. The components $\sigma_{1,2} (\omega)$ of the complex conductivity were inferred from the expressions (8) of~\cite{Drichko_PhysRevB.62.7470}.

\section{Results and discussion}
\begin{figure}[h]
\centering
\includegraphics[width=8.5cm]{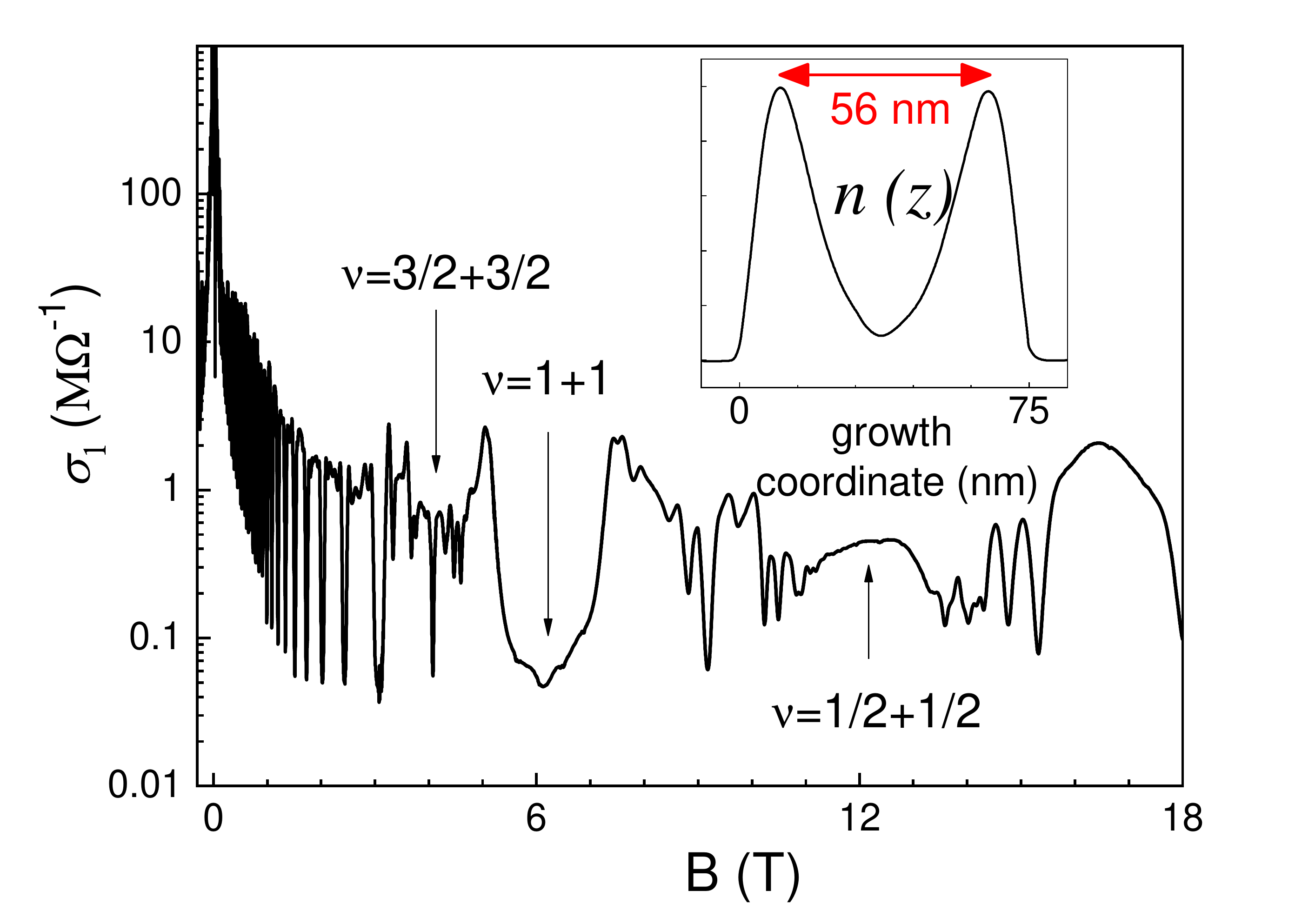}
\caption{Magnetic field dependence \ild{of the real part of AC conductance}, $\sigma_1 (\omega)$, at $T=20$~mK and $f=30$~MHz. The dependences $\sigma_1 (B)$ for other frequencies in the domain studied look similarly. Inset: Charge distribution in the QW.
\ild{
\ymg{Sums in the filling factors}
demonstrate that the oscillations are arising from two layers with close concentration $\sim 1.5 \times 10^{11} $~cm$^{-2}$.}
\label{fig1}}
\end{figure}

%
Shown in Fig.~\ref{fig1} the magnetic field dependence of $\sigma_1$ at $T=20$~mK and $f\equiv \omega/2\pi =30$~MHz. The inset demonstrates the charge distribution inside the QW.

\ymg{Before starting discussion let us comment on}
\ild{
the relationship between electron density and the filling factors in our system. }
\ymg{Conventionally,}
\ild{
	in a bi-layer structure formed in a wide QW,  the total electron density is determined from the position of the minimum (maximum in the case of $\nu$=1/2) of the FQHE resistance}
\ymg{(conductance)}
\ild{
 oscillations in strong magnetic fields. }
\ymg{The difference between}
\ild{
the charge densities in the layers is derived using Fourier analysis of magnetoresistance in the fields $B < 0.4-0.5$~T.

However, the case of a bi-layer sample}
\ymg{signal from which  shown in}
\ild{
 Fig.~\ref{fig1} requires a special }
\ymg{consideration.}
\ild{
The oscillations pattern
}
\ymg{close to}
\ild{$B=$12.35~T looks }
\ymg{almost identical to that observed  in single layered structures close to $\nu=1/2$, but
for a twice smaller concentration $\sim 1.5 \times 10^{11} $~cm$^{-2}$. Accordingly, for total concentration $3 \times 10^{11} $~cm$^{-2}$ such a pattern should be located at $B \approx 25$~T.}
%
\ild{
We believe that in high  magnetic fields, the layers in a }
\ymg{wide QW}
\ild{
are practically independent (especially at so small $\Delta_{SAS}=0.42$~meV as in our sample), and the conductances from different layers with almost the same concentration are simply added.
}
\ild{
The fact that oscillations at $B=$12.35~T corresponds to $\nu$=1/2 for each layer with a concentration $\sim 1.5 \times 10^{11} $~cm$^{-2}$ is emphasized in Fig.~\ref{fig1}
}
\ymg{presentation of the filling factors as sums of the layers' contributions.
}

\begin{figure}[h]
\centering
\includegraphics[width=8.5cm]{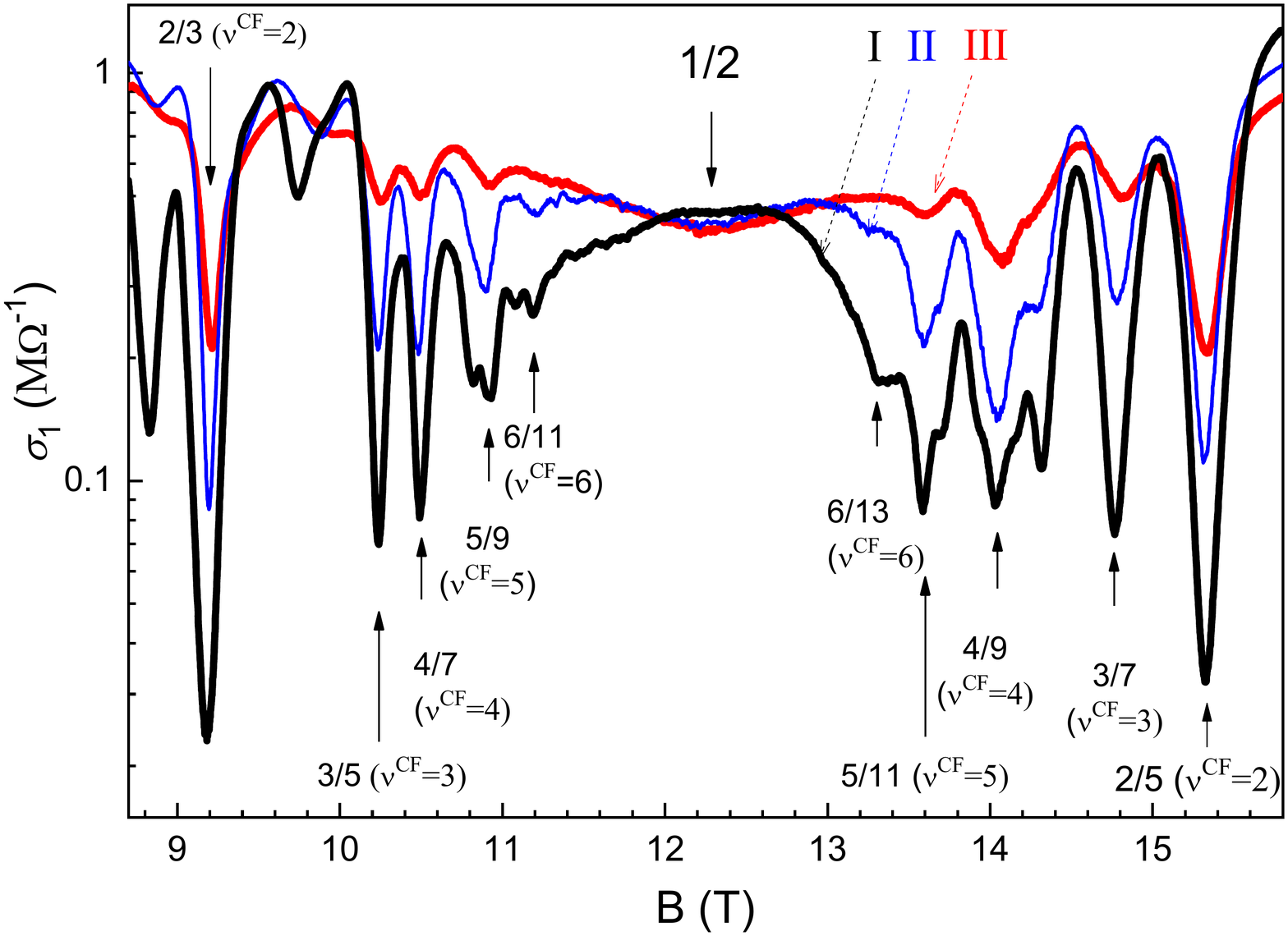}
\caption{Magnetic field dependences \ild{of the real part of AC conductance, $\sigma_1 (\omega),$}
 for different temperatures 20-480~mK zoomed in the vicinity of $\nu = 1/2$. Arrows represent the values of $\nu$ calculated for the density $n_e=1.5 \times 10^{11}$~cm$^{-2}$. $T$,~mK: I - 20, II - 265, III - 480.
The values of the filling factor for \textit{composite fermions}, $\nu^{CF}$, calculated from Eq.~(\ref{eq01}) are shown in the parentheses.
\label{fig2}}
\end{figure}

The vicinity of the filling factor $\nu =1/2$ is zoomed in Fig.~\ref{fig2}. Arrows represent the values of $\nu$ calculated for the density $n_e= 1.5 \times 10^{11}$~cm$^{-2}$. The values of the filling factor for \textit{composite fermions}, $\nu^{CF}$, are shown in the parentheses. They are calculated from the expression
\begin{equation} \label{eq01}
\nu = \nu^{CF}/(2 \nu^{CF} \pm 1 ).
\end{equation}
The picture looks as a typical one for composite fermions, which is usually observed in relatively narrow QWs, see, e.g., \cite{Willett_PhysRevB.47.7344,Goldman_PhysRevLett.72.2065}

\begin{figure}[h!]
\centering
\includegraphics[width=8.5cm]{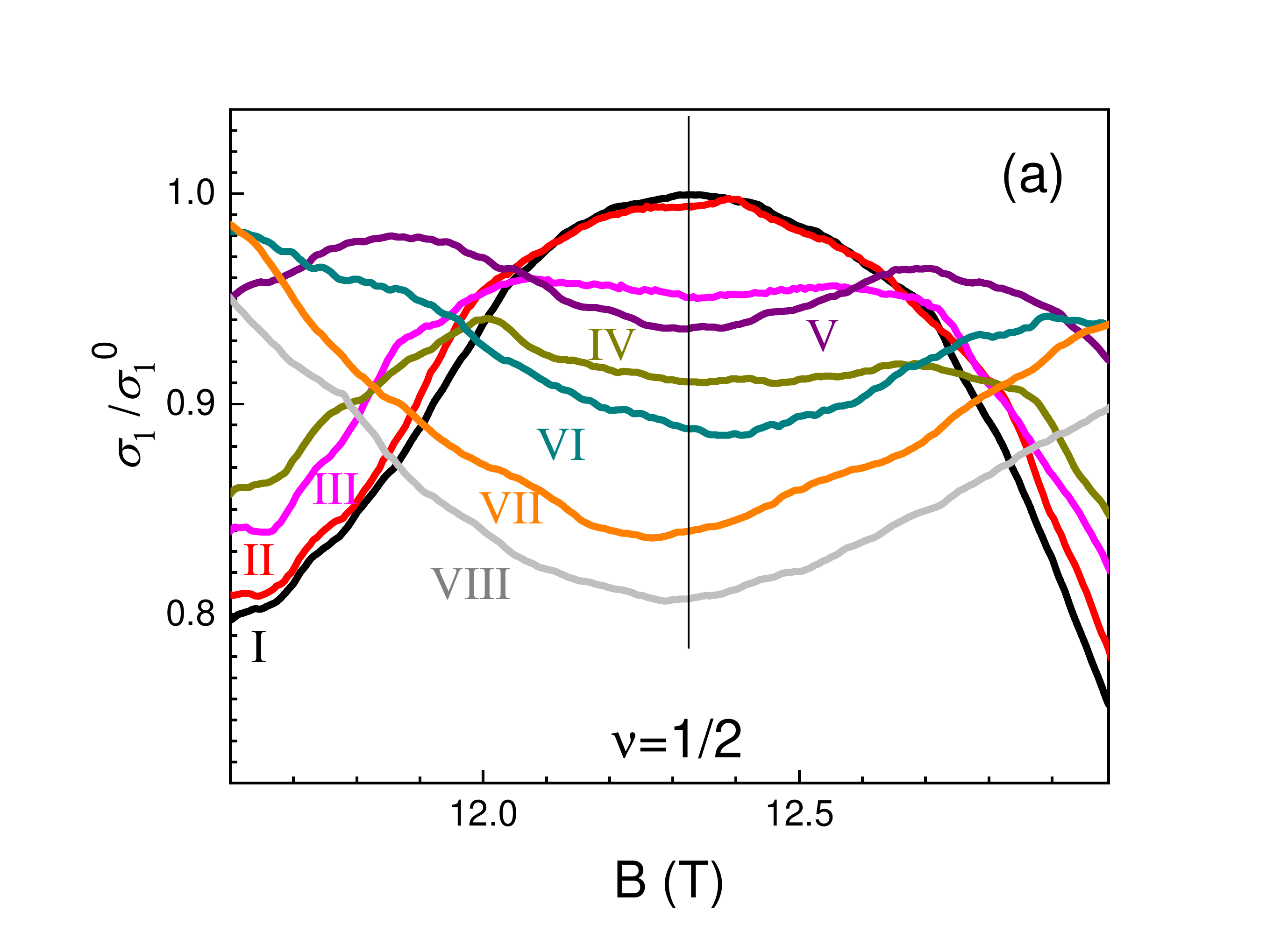} \\
\includegraphics[width=8.5cm]{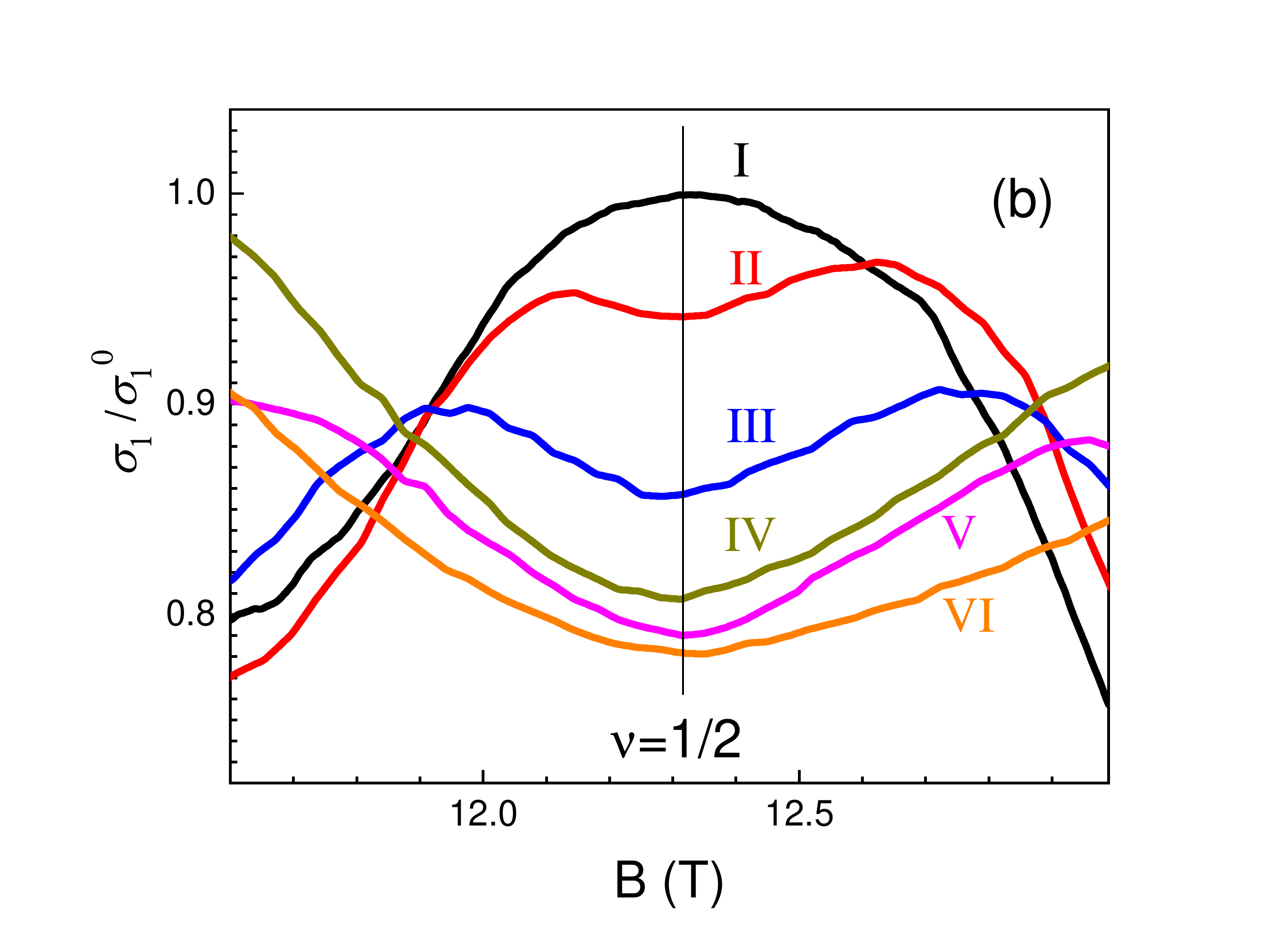}
\caption{Upper panel: \ild{Reduced conductance $\sigma_1/\sigma_1^0$} at $f=30$ MHz  versus magnetic field
for different temperatures. $T$ (mK): I - 20, II - 40, III - 110, IV - 155, V - 210, VI - 280, VII - 400,  VIII - 520. Lower panel: $\sigma_1$ at $f=30$ MHz  versus magnetic field for different SAW powers.
The SAW power introduced into the sample (W): I: - 3.5$\times$10$^{-10}$, II: - 3.4$\times$10$^{-9}$, III: - 3.4$\times$10$^{-8}$, IV: - 3.4$\times$10$^{-7}$, V: - 3.5$\times$10$^{-6}$; VI: - 3.5$\times$10$^{-5}$.
\label{fig3-4}}
\end{figure}

The magnetic field dependences \ild{of conductance} in a close vicinity of $\nu=1/2$ for different temperatures and SAW powers are further analyzed in Fig.~\ref{fig3-4}. The upper panel shows the dependences \ild{of reduced conductance $\sigma_1/\sigma_1^0$ ($\sigma_1^0$ is the value of $\sigma_1$ at $T$=20~mK, $\nu=$~1/2 measured at lowest used SAW power)} for $f=30$~MHz, different temperatures and the lowest used SAW power.
At the lowest temperature (20~mK) $\sigma_1$ has a maximum at $\nu=1/2$, which crosses over to a minimum at $T > 400$~mK.

Dependence \ild{of conductance} on the power (Fig.~\ref{fig3-4}, lower panel) also shows a crossover in the magnetic field dependence from a maximum to a minimum as the SAW power increases. One notices that increase of the power manifests itself similarly as an increase in the temperature.

Note that the behavior mentioned is observed in the entire frequency domain studied. It is worth mentioning that the values of $\sigma_1$ in the extrema are essentially frequency independent. In addition, within the entire frequency domain studied $\sigma_1 \gg \sigma_2$.

\begin{figure}[h!]
\centering
\includegraphics[width=8.5cm]{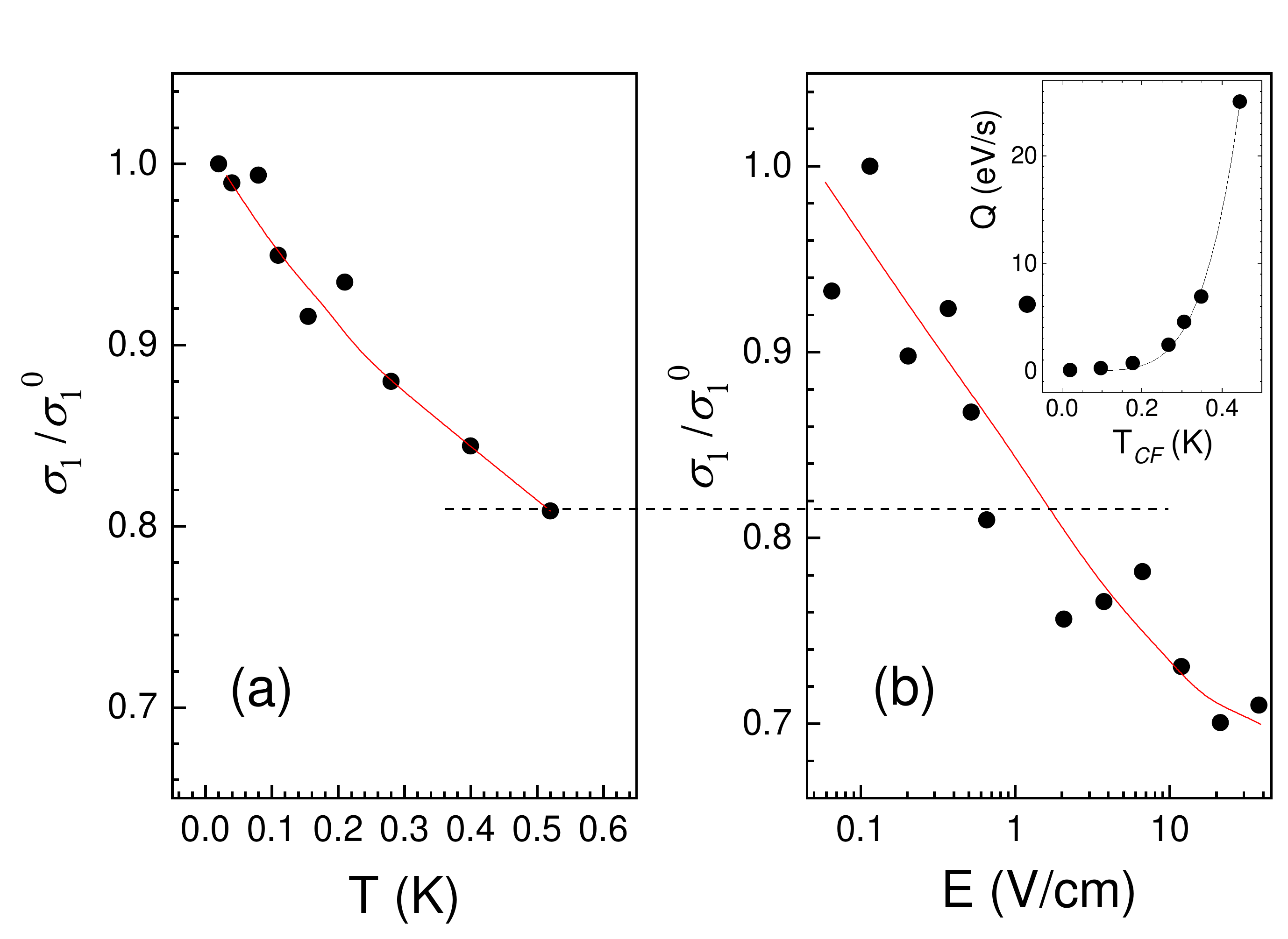}
\caption{ Dependences of \ild{reduced} conductance $\sigma_1/\sigma_1^0$ on the temperature (left panel) and on the SAW electric field  (right panel), both at $\nu = 1/2$. $f=30$~MHz. The temperature dependence of $\sigma_1$ is measured in the linear regime; the dependence $\sigma_1$ on the SAW electric field is measured at $T=20$~mK. \ild{Inset:  The energy losses rate per electron vs $T_{CF}$. Line is the result of fitting the experiment with $Q=A_5 (T_{CF}^5-T^5_{lattice})$.}
\label{fig5}}
\end{figure}

Shown in Fig.~\ref{fig5} are the dependences of $\sigma_1$ on the temperature (left panel) and on the SAW power (right panel), both at $\nu = 1/2$.
The temperature dependence of $\sigma_1$ is relatively weak which is typical for a compressible state of composite fermions~\cite{HLR_PhysRevB.47.7312}. This conclusion is compatible with very weak frequency dependence of $\sigma_1$ which is characteristic for acoustic properties of the gas of CFs at $k\ell \ll 1$. Here $k = \omega/V$ is the SAW wave vector while $\ell$ is the mean free path of the CFs.

The estimates along Ref.~\cite{HLR_PhysRevB.47.7312} show that this indeed the case for our structure and the frequency domain studied. Indeed, at $\nu = 1/2$ assuming that the electrons are fully spin-polarized\footnote{\ymg{Assumption of full spin polarization is not crucial since the condition $k\ell \ll 1$ is fulfilled event if the spin polarization is incomplete.}}
 we get
\begin{equation}
B_{1/2}=4\pi \hbar c n_e/e, \
l^2_{B} (1/2)=
c\hbar/eB_{1/2}
\to k_F = 1/l_B.
\end{equation}
Putting $B_{1/2}=12.3$~T we estimate the electronic Fermi momentum as $k_F \approx 1.4\times 10^6$~cm$^{-1}$. The SAW wave vector, $k$, can be estimated as $k =\omega /V$. For
$\omega/2\pi = 10^8$ Hz and sound velocity $V=3 \times 10^5$~cm/s we have
$k\approx  2\times 10^3$~cm$^{-1}$. Therefore, $k \ll k_F$.

The transport relaxation time of CFs can be estimated from Eq.~(5.7) of \cite{HLR_PhysRevB.47.7312}
\begin{equation}
\frac{1}{\tau_{\textrm{tr}}}=\frac{4\pi \hbar n_{\textrm{imp}}}{m^*k_Fd_s}=\frac{n_{\textrm{imp}}}{n_e} \frac{\hbar k_F^2}{m^* (k_Fd_s)}.
\end{equation}
Here $m^*$ is the effective mass of a CF; according to~\cite{HLR_PhysRevB.47.7312} it can be estimated $4m_b$ where $m_b$ is the electron effective mass; $d_s$ is the spacer value.
\ymg{Unfortunately, the values of $m^*$ reported in different papers are significantly different. Therefore,  we do not use the above expression for finding $n_\textrm{imp}$.
 }
 \ild{The structure under study was designed with  and spacer $d_s=$820$~{\AA}$ providing near $\nu=$1/2 $n_e=1.5 \times 10^{11}$~cm$^{-2}$ as explained earlier.}
%
\ymg{Using the above values} we get:
\ymg{
\begin{multline}
\ell \equiv \frac{\hbar k_F \tau_{\textrm{tr}}}{m^*}=\frac{n_e}{n_\textrm{imp}}\frac{\hbar k_F }{m^*}\frac{m^*(k_Fd_s)}{\hbar  k_F^2} =0.1 d_s \approx \nonumber \\ \approx 8 \times 10^{-7}~\textrm{cm},
k_F=1.4 \times 10^{6}~\textrm{cm}^{-1}; \   k_F d_s \approx 11.5.
\end{multline}
}
We conclude that in all our frequency domain the inequality $k\ell \ll 1$ is met. It implies that  $\sigma_2 (\omega) \ll \sigma_1(\omega)$ and the AC conductance practically coincides with the DC conductance.
\ymg{Note that this conclusion remains valid for all realistic values of $m^*$.}

\ild{
Figure~\ref{fig5} demonstrates dependences of reduced conductivity $\sigma_1 / \sigma_1^0$
 at $f=30$~MHz
and
 $\nu$=1/2 on (a) temperature measured in the linear regime and
\ymg{(b) in the AC electric field $\mathbf{E}$  produced by the SAW and measured at $T=20$~mK.\footnote{Relationship between the field amplitude $E$ and the intensity $W$ of the SAW was analyzed
in~\cite{HeatSAW}.}  Similarity of the curves leads to a conclusion that the observed dependence $\sigma_1(E)$ is associated with heating of the CFs.}
%
 %
The latter may be described by
introducing
\ymg{an effective temperature $T_{CF} >T_{lattice}$ of composite fermions
which }
can be determined by comparing the curves of Fig.~\ref{fig5}(a) and Fig.~\ref{fig5}(b).
So we can establish a correspondence between the temperature of
the CF and the SAW electric field. Then we can determine the absolute energy loss rate $Q=e \mu E^2$ per
electron.
\ymg{The variation of $Q$ with $(T_{CF}^S-T^S_{lattice})$ obtained from the experimental
results is shown in  inset of Fig.~\ref{fig5}(b) and could be well fitted
with $S$=5.}

Analysis of the dependence $Q$ on effective temperature  allows }
\ymg{one to judge on
the mechanism of the energy relaxation of composite fermions.  The fitted parameter $S=5$ for \textit{electrons} would indicate on electron scattering from the piezoelectric
potential of the acoustic phonons (PA-scattering) with strong screening~\cite{Ridley}. Unfortunately, we are not aware of relevant theory of the energy relaxation of CFs.
}

What remains to be explained is the inversion of the oscillation with increase of either temperature, or the SAW power.  Such an inversion was observed, e.g., in GaAs double quantum wells with two occupied size-quantization levels and attributed to formation of essentially non-equilibrium distribution function, see, e.g.,~\cite{Bykov_JETP_Lett} and references therein.  However, this explanation would be valid for relatively weak magnetic fields, $B \ll B_{1/2}$ where $B_{1/2}$ corresponds to $\nu = 1/2$.
Note that the observed temperature dependence of $\sigma_1$ is qualitatively similar to  that observed in~\cite{Willett_PhysRevB.47.7344,Rokhinson_PhysRevB.52.R11588}. In that paper the complex conductance in the frequency domain $0.9-3.4$~GHz (corresponding to $k\ell \gg 1$) was measured in \ild{the high mobility single interface AlGaAs/GaAs with concentration of $n_e=1.5 \times 10^{11}$~cm$^{-2}$}.

Though our results do not allow unique identification of the conduction mechanism close to $\nu = 1/2$ one can notice that both the dependences on temperature and magnetic field are compatible with the behavior of a gas of composite fermions.
\ymg{Indeed, as it  follows from  Eq.~(6.45) of
Ref.~\cite {HLR_PhysRevB.47.7312}, close to $\nu = 1/2$ $\rho_{xx}=\rho_{xx}^{CF}$. Taking into account that in this region $\sigma_{xy}^{CF} \to 0$ and therefore  $\rho_{xx}^{CF}=1/\sigma_{xx}^{CF}$
 one finds that~\cite{Rokhinson_PhysRevB.52.R11588}
}
%
close to $\nu = 1/2$ the conductivity tensors of CFs and electrons are related as
\begin{equation}\label{eq:05}
\sigma_{xx}^{CF} = \frac{\sigma_{xx}^2 +\sigma_{xy}^2 }{\sigma_{xx}}\approx \frac{\sigma_{xy}^2 }{\sigma_{xx}}
\end{equation}
\cred{(see also~\cite{Rokhinson_PhysRevB.52.R11588}).}
Therefore, the observed weak temperature dependence of $\sigma_{xx}$ at $\nu=1/2$ can be interpreted as originated from a quantum contribution to
$\sigma_{xx}^{CF}$.\footnote{A similar reasoning was used in~\cite{Rokhinson_PhysRevB.52.R11588}.}
The low-temperature maximum of $\sigma_{xx}(\Delta B)$ at $\Delta B \equiv B - B_{1/2} =0$ is qualitatively compatible with the quantum negative magnetoresistnce of composite fermions; \cred{note that according to Eq.~(\ref{eq:05})
$\sigma_{xx}\propto 1/\sigma_{xx}^{CF}$.} Increase of temperature may lead to its suppression, and CFs show classical positive magnetoresistance. As a result, the maximum in the $\sigma_{xx}(\Delta B)$ dependence crosses over to a minimum. Unfortunately, we are not aware of a quantitative theory of the quantum contribution to $\sigma_{xx}^{CF}$, which would be valid for a quasi-bi-layer structure.
\cred{In addition, the observed deviations seem to be too large to be interpreted as quantum contributions.}
Therefore, we restrict ourselves with
\cred{purely}
qualitative considerations.

\begin{figure}[h!]
\centering
\includegraphics[width=8.5cm]{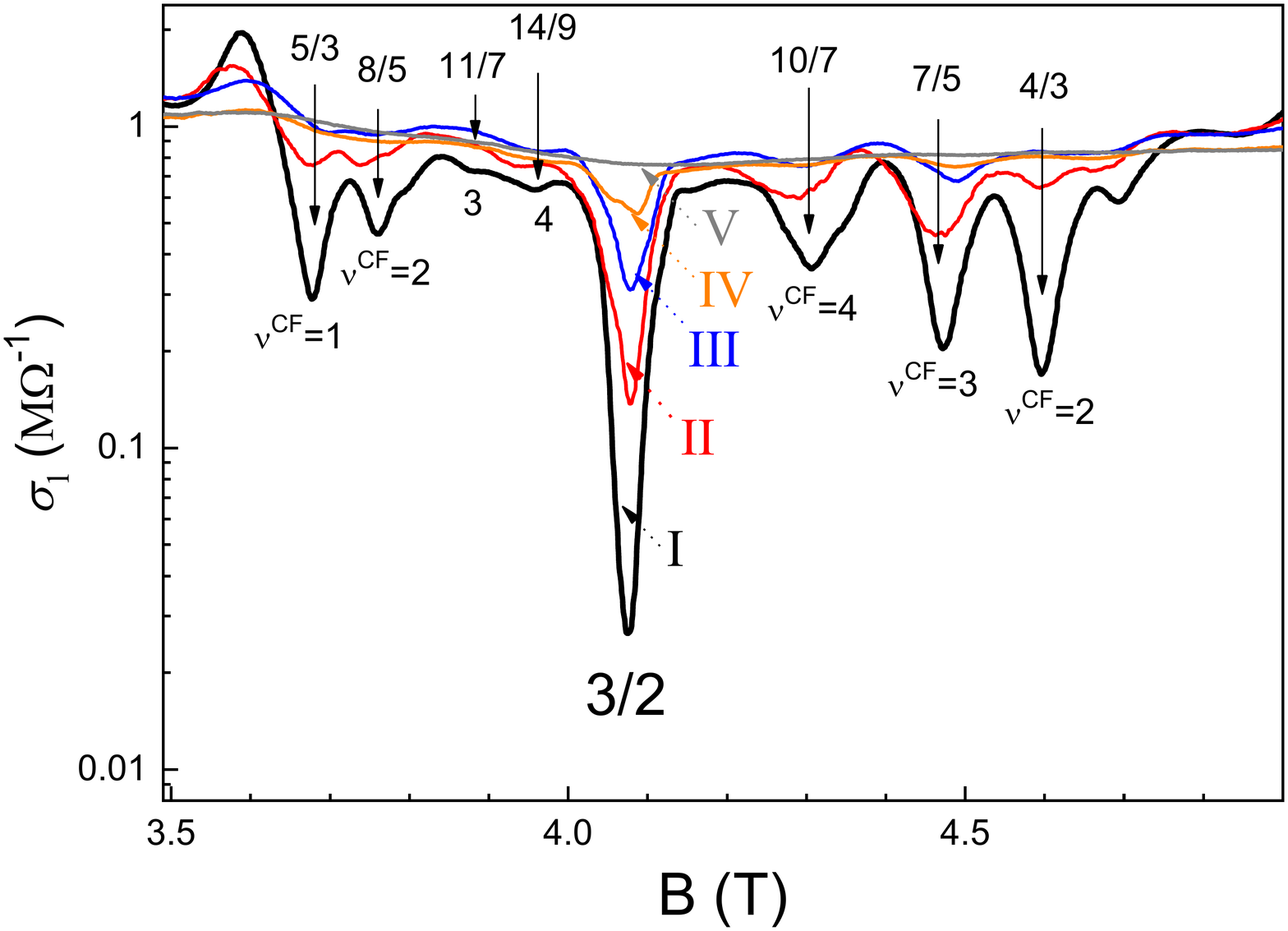}
\caption{ Dependence of $\sigma_1$ on magnetic field in the vicinity of $\nu = 3/2$ for different temperatures $T$,~mK: I -- 20, II -- 107, III -- 200, IV -- 310, V -- 520. $f=85$~MHz.
Arrows represent the values of $\nu$.
Also shown the values of the filling factor for \textit{composite fermions}, $\nu^{CF}$.
\label{fig6}}
\end{figure}

\begin{figure}[h!]
\centering
\includegraphics[width=8.5cm]{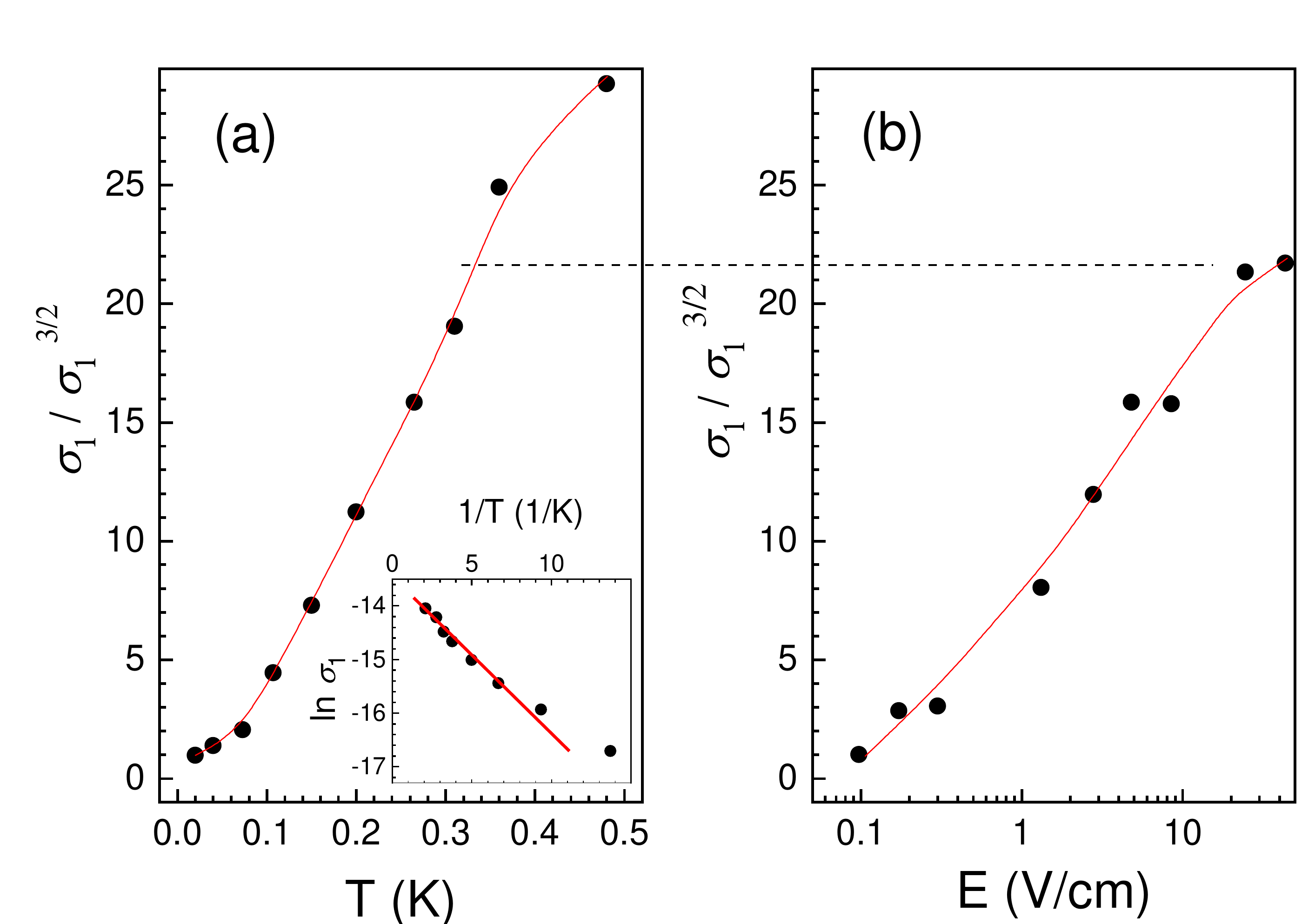}
\caption{ Dependences of \ild{reduced conductance $\sigma_1 / \sigma_1^{3/2}$ on the temperature (left panel) and on the SAW electric field  (right panel), both at $\nu = 3/2$. $f=85$~MHz ($\sigma_1^{3/2}$  is the value of $\sigma_1$ at $T$=20~mK, $\nu=$~3/2, measured at lowest used SAW power)}. \ild{The inset: Arrhenius plot evidencing an activation character of conductivity $\sigma_1 \propto \exp (-\Delta_a/2 T)$ in the temperature range 100-500 mK, $\Delta_a=(50 \pm 3)$~mK.} The temperature dependence is measured in the linear regime; the dependence on the SAW power is measured at $T=20$~mK.
\label{fig7}}
\end{figure}

It is worth mentioning that the behavior of $\sigma_1$ close to $\nu =3/2$ does not correspond to a compressible state with CFs. It is rather corresponds to a typical picture of incompressible state showing the FQHE. This is illustrated in Fig.~\ref{fig6} where the magnetic field dependences of $\sigma_1$ are shown.  The deep minimum at $\nu$=3/2 is surrounded by two series of oscillations, corresponding to the composite fermion states with $\nu^{CF}$, determined
from the equations
\begin{equation}
\nu=2-\frac{\nu^{CF}}{2\nu^{CF}\mp 1}
\end{equation}
for the right and left series, respectively.
Note that at $\nu =3/2$ the AC conductance $\sigma_1$ exponentially increases with increasing temperature. Such a dependence (Fig.~\ref{fig7}a) is characteristic for  the localized charge carriers. Dependence of $\sigma_1$ on the SAW electric field (Fig.~\ref{fig7}b) is analogous to its temperature dependence.

\paragraph{Conclusion} --
We conclude that  it is a compressible state with composite fermions that is realized in our sample close to $\nu =1/2$. The conclusion is based on
\cred{
observed weak metal-type dependences of $\sigma_1(\omega)$ on both temperature and frequency, as well as on the similarity of dependences on temperature and SAW power.
}
 These behaviors essentially differ from those close to $\nu = 3/2$, where a typical picture of in-compressible state showing the FQHE. The compressible state that we've observed in our sample seems to be unusual.
\ymg{Indeed, our observation of a compressible state at $\nu = 1/2$ contradicts to previous studies. For example, according to $\Delta_{SAS}$ – $n_e$ and $W$  – $n_e$ diagrams reported in~\cite{Shabani_PhysRevB.88.245413}, where $W$ is a QW width (see Figs. 5 and 6 in~\cite{Shabani_PhysRevB.88.245413}, respectively) a QW with our parameters at $\nu = 1/2$ should be insulating or showing FQHE. Also, properties of our QW also cannot be described by the dependences obtained in~\cite{Manoharan_Shayegan_1996}. Further investigations are required to understand the cause of disagreement between our results and those reported by others.
}

\paragraph{Acknowledgments} --
The authors would like to thank E. Palm, T. Murphy, J.-H. Park,
A. Bangura, and G. Jones for technical assistance. Partial support from
program No.~1 ``Physics and technology of nanostructures, nanoelectronics and diagnostics'' of the Presidium of RAS (I.~Y.~S.)
 and the Russian Foundation for Basic Research projects
19-02-00124 (I.~L.~D.)
is gratefully acknowledged. The National High Magnetic Field Laboratory is supported by the National Science Foundation through NSF/DMR-1157490 and NSF/DMR-1644779 and the State of Florida. The work at Princeton was supported by Gordon and Betty Moore Foundation through the EPiQS initiative Grant GBMF4420, and by the NSF MRSEC Grant DMR-1420541.

\bibliography{mybibfile}

\end{document}